\documentstyle[12pt]{article}

\begin{document}
\setcounter{page}{0}
\begin{titlepage}
\title{Fermionic Ising Glasses with BCS Pairing Interaction.Tricritical 
   Behaviour}
\author{ S. G. Magalh\~aes\\
	 Departamento de Matematica,
	 Universidade Federal de Santa Maria,\\
	 97119-900 Santa Maria, RS, Brazil.\\
	 Alba Theumann\\
	 Instituto de F\'{\i}sica\\ 
	 Universidade Federal do 
	 Rio Grande do Sul,\\
	 Av.Bento Gon\c{c}alves 9500,
	 C.P. 15051\\
	 91501--970 Porto Alegre, RS, Brazil }
\date{}
\maketitle
\thispagestyle{empty}
\begin{abstract}
\normalsize
\noindent
We have examined the role of the BCS pairing mechanism in the 
formation of the magnetic moment and henceforth a spin glass (SG) 
phase by studying a fermionic Sherrington-Kirkpatrick model with a 
local BCS coupling between the fermions. This model is obtained 
by using perturbation theory to trace out 
the conduction electrons degrees of freedom in conventional superconducting alloys.
The model is formulated in the path integral formalism where the spin 
operators are represented by  bilinear combinations of Grassmann fields
and it reduces to a single site problem that can be solved within the
static approximation with a replica symmetric Ansatz. We argue that 
this is a valid procedure
for  values of temperature above the de Almeida-Thouless instability line.
The phase diagram in the T-g plane,
where g is the strength of the pairing interaction, for fixed
variance $J^2/N$ of the random couplings $J_{ij}$, exhibits
three regions: a normal paramagnetic (NP) phase, a spin glass (SG)
phase and a pairing (PAIR) phase where there is formation of
local pairs.The NP and PAIR phases are separated by a second
order transition line $g=g_{c}(T)$ that ends at a tricritical point
$T_{3}=0.9807J$, $g_{3}=5,8843J$, from where it becomes a first order 
transition line that meets the line of
second order transitions at $T_{c}=0.9570J$ that separates 
the NP and the SG phases. For $T<T_{c}$ the SG phase is
separated from the PAIR phase by a line of first order transitions.
 These results agree 
qualitatively with experimental data in $\mbox{Gd}_{x}\mbox{Th}_{1-x}\mbox{RU}_{2}$.
\end{abstract}
\vspace*{0.5cm}
\newline 
\noindent
\end{titlepage}
\setcounter{page}{1}
\section*{1. Introduction}

Experimental evidence in cuprate superconductors \cite{1} 
exhibit a very rich phase diagram that includes structural, 
antiferromagnetic, insulator-metal, superconducting and spin glass 
transitions, that depend strongly on the dopant concentration. 
The coexistence of spin glass ordering and superconductivity has been also 
observed in conventional superconductors doped with magnetic 
impurities \cite{2}.
 Theoretical studies of conventional spin glass superconductors have
 focused in calculations
 of the superconducting density of states in the presence of localized 
magnetically ordered impurities.
 These systems are well described \cite{3} by a Hamiltonian where the
superconducting electrons are represented
by a conventional BCS Hamiltonian and they interact  with the localized 
magnetic impurities via the s-d exchange interaction. Theoretical studies of 
superconductive glass 
models that describe random arrays of Josephson junctions have been performed 
both for classical \cite{4} 
and recently in a quantum model \cite{5}. 

Our motivation in this paper is to study the interplay of the mechanisms 
that lead to spin glass ordering and BCS pair formation in  
a fermionic Ising spin glass model with BCS pairing 
among localized fermions of opposite spins. We argue in the Appendix that this effective 
Hamiltonian is obtained from the model of Ref. \cite{3} by tracing out the degrees 
of freedom of the superconducting electrons to second order in the s-d exchange interaction,
 when the localized spin operators are represented by bilinear combinations of fermions. 
In this case, besides the known RKKY interaction between localized
spins we obtain an exchange induced pairing interaction between localized fermions,
 mediated by the 
superconducting electrons. This model allow us to investigate the competitions 
between frustration 
and double occupation of the sites in a half-filling situation.

Since the introduction of the Sherrington-Kirkpatrick \cite{6} 
(SK) model to describe infinite-ranged Ising spin glasses, a vast 
amount of work was devoted to the study of analogous quantum spin glass 
(QSG) models with different and interesting low temperatures properties.

In an early seminal paper \cite{7}, Bray and Moore used Feynman's
functional integrals formalism with a fictitious time $ 0<\tau < \beta$,
$\beta =\frac{1}{T} $, to analyze the quantum Heisenberg spin glass model.
By using the static approximation to evaluate the spin-spin correlation
functions, they established the existence of a phase transition at 
finite temperatures. 
This formalism has been extended recently to the study of quantum 
fluctuations in  related spin glass models \cite{8,9,10}.
 The authors in Ref.\cite{8} and Ref.\cite{9} report on unconventional 
time(frequency) behaviour of the correlation functions at $T=0$.
 In a remarkable later
work \cite{10}  Grempel and Rozenberg found the exact numerical solution 
of Bray and Moore's equations \cite{7} for $S=1/2$, and they demonstrate the 
existence of an ordered spin glass phase below  a finite critical temperature .
Also the spin-spin correlation function $Q(\tau)$ is found to be roughly constant
and equal to its classical value within a range of temperatures around the critical 
point, what it seems to justify the use of the static Ansatz of Ref.\cite{7}
at not very low temperatures in the Heisenberg spin glass.\\
Other functional integral techniques have been used earlier 
to study QSG models where 
the spin operators are represented by bilinear combinations of fermionic 
(anticommuting) Grassmann fields, both in the anisotropic 
(Ising) \cite{11}  and the isotropic (Heisenberg) \cite{12}  limits 
within a replica symmetric (RS) theory. 

The static approximation was used, and it turned out to be exact,
 in the fermionic Ising model,
while the fermionic Heisenberg model was solved by combining
the static approximation for the order parameter with an instantaneous 
approximation for the retarded susceptibility.

Recent work \cite{13} 
demonstrated the existence of several characteristic temperatures in 
both models, with the de Almeida-Thouless \cite{14} instability occurring 
at a temperature $T_ {1}$ lower than the spin glass transition $T_{SG}$. 
In the isotropic fermionic model \cite{12} there exists still a lower 
temperature $T_{2}<T_{1}$, at which the replica symmetry 
stability is restored. The region of RS instability is characterized 
by a negative entropy in the anisotropic \cite{11} fermionic model, 
while the entropy remains positive in the isotropic model \cite{12} 
but the specific heat changes sign in the RS instability region.\\
 
The anisotropic (Ising) QSG model \cite{11, 15}  deserves some special 
discussion. In this particular case, the spin operator $S^{z}_{i}$ 
commutes with the particle number operator $n_{is}=0$ or $1$, 
and thus it would not be necessary to employ the functional integral 
formulation since the Hamiltonian is diagonal in occupation number 
operators. However, there still remains an important difference 
between the fermionic and classical SK spin glass: in the quantum case 
the diagonal component of the order parameter in replica space is no 
longer constrained to unity. Consequently, the susceptility $\chi$ 
emerges in the problem with an important new role and the spin glass 
order parameter has to be determined coupled to $\chi$. By adding to the 
fermionic Ising \cite{11,13} a term that favors BCS pairing, the use of functional 
integrals becomes necessary as the Hamiltonian does not commute with the particle
 occupation number operators. 

There is a crucial aspect that characterizes the representation of 
spin operators in Fock space, because there are four quantum states 
at every site \cite{12} , two of them non-magnetic, and the quantum 
statistics that controls number occupation can induce unusual phase 
transitions.
In other words,  the QSG frustration can be disrupted as 
long as we have access with equal probability to the magnetic and 
the non-magnetic states at each site. In fact, a recent 
paper \cite{16} has reported tricritical behaviour  in the fermionic 
Ising QSG model, within the static approximation,
 by varying the electronic concentration. 
This raises the question if the effects that come from the relative 
occupation of magnetic and non-magnetic states can be properly 
exploited, and consequently to produce unusual phase transitions 
even if the average occupation per site is kept constant and equal to one.
The main difference between Ref. \cite{16} and ours resides in the mechanism 
that controls the magnetic moment formation on the sites. 
They achieve that by 
varying the electronic concentration while we have a pairing 
mechanism that energetically favors the double occupation and therefore non-magnetic states in the half-filling situation.

This paper is structured as follows: in Sec.2 we study the 
model derived in the Appendix and find the thermodynamic potential, together with the 
saddle point equations for the order parameters.
 In Sec.3 we 
discuss the nature of the phase transitions in the resulting phase 
diagram in the T-g plane, where g is the strength of the pairing 
interaction, for fixed variance of the random couplings $J_{ij}$. 
Lowering the temperature, for high values of g, there is a line of 
second order transitions from normal to the pair formation phase, 
that ends at a tricritical point $T=T_{3}$ and $g=g_{3}$, 
characterized by the simultaneous vanishing of the  two first 
coefficients in the Landau expansion of the free energy \cite{17}. 
From there on, the line becomes one of first order transitions until 
it meets the line of second order spin glass transitions at 
$T=T_{c}$. For $T<T_{c}$ the first order transition line separates 
the spin glass and pair formation phases. 
All these results were obtained by using the static approximation. 
As we discussed previously, we expect this to be a justifiable Ansatz
because all the relevant temperatures are of the order $ \beta J\approx 1$
\cite{7} , and our theory is not applicable to very low temperatures due to the
de Almeida-Thouless instability.\\
We reserve Sec.4 for 
discussions and comparison with other models and the experimental data \cite{2}.    
\section*{2. General Formulation}

Conventional spin glass superconductors are well represented by a Hamiltonian where
 the conduction electrons are described by a BCS Hamiltonian and they interact via an
 effective s-d exchange term with randomly localized magnetic impurities \cite{3}.
 We show in the Appendix that, when the localized spins are 
represented in terms of fermions,
the degrees of freedom of the superconducting electrons 
can be integrated using second order perturbation theory
in the exchange interaction $J_{sd}$ to give rise 
to an effective BCS pairing interaction among the fermions, 
besides the very well known RKKY 
interaction among the localized spins.
In the mean field spirit we are lead to consider the following Hamiltonian:
\begin{eqnarray}
 \overline{H}= H-\mu N= -\sum_{ij}J_{ij}S^{z}_{i}S^{z}_{j}- 
\mu\sum_{i}\sum_{s=\uparrow,\downarrow}c^{\dagger}_{is}c_{is}-\nonumber\\
-\frac{g}{N}\;\sum_{i,j}c^{\dagger}_{i\uparrow}c^{\dagger}_{i\downarrow}
c_{j\downarrow}c_{j\uparrow}
\label{2.1}
\end{eqnarray}
where the operator $S^{z}_{i}$ is defined as
\begin{equation}
S^{z}_{i}= c^{\dagger}_{i\uparrow}c_{i\uparrow}-
c^{\dagger}_{i\downarrow}c_{i\downarrow}\;,
\label{2.2}
\end{equation}
$\mu$ is the chemical potential, $c^{\dagger}_{is}$ ($c_{is}$) are 
fermions creation (destruction) operators and $s=\uparrow$ 
or $\downarrow$ indicates the spin projection.
The coupling  $J_{ij}$ is an independent random variable with the 
distribution
\begin{equation}
P(J_{ij})=e^{-J_{ij}^{2}N/2J^{2}}{\sqrt{N/2\pi J^{2}}}\;.
\label{2.3}
\end{equation}
The first two terms in the Hamiltonian of equation (\ref{2.1}) describe 
a fermionic Ising spin glass \cite{11,13,15} while the last term is 
a BCS-like pairing interaction and corresponds to the mechanism that 
favors the double occupation of sites \cite{18}.

Our ultimate goal is to reduce this problem to a one-site problem. 
Functional integration techniques have proved to be a suitable approach 
for disordered quantum-mechanical many-site problems, as it has been for 
classical problems \cite{19}. 
Furthermore, this formulation showed to be quite successful to 
describe the usual superconductive transition with a BCS coupling \cite{20} 
and in the presence of transition metal impurities \cite{21}. In that case, the 
particle-hole transformation introduced by the use of Nambu matrices 
within the static approximation made the problem with a BCS coupling 
solvable, because it becomes a mean field theory 
in momentum and frequency space .
As the static approximation is exact for the
BCS problem and is also exact for the fermionic Ising glass
\cite{11,12}, we expect it to give reliable interpolation results  here
for finite temperatures \cite{10}.  Our theory is not valid 
at very low temperatures due to the de Almeida-Thouless instability,
then we are not concerned with the singular behaviour
found at $T= 0$ in other models \cite{8,9}.

In the Lagrangian formulation \cite{11,19,21} the 
partition function is expressed as
\begin{equation}
Z=\int D(\phi^{\ast}\phi) e^{A}
\label{2.4}
\end{equation}
where the action A is given by
\begin{equation}
A=\int_{0}^{\beta}[\sum_{is}\phi^{\ast}_{is} (\tau)\frac{d}{d\tau}
\phi_{is}(\tau)-\overline{H}(\phi^{\ast}_{is}(\tau),
\phi^{\ast}_{js}(\tau))]d\tau\;.
\label{2.5}
\end{equation}
In both expressions $\phi^{\ast}_{is}(\tau)$ and $\phi_{is}(\tau)$ are 
anticommuting Grassmann variables, $\tau$ is a complex time and 
$\beta$ the inverse absolute temperature.

In order to apply the particle-hole transformation within the 
static approximation and to make explicit our central approximation, 
we work with time Fourier transformed quantities. 
Therefore, the pairing part of the action becomes
\begin{equation}
A_{pairing}=\frac{\beta g}{N} \sum_{\Omega}\sum_{ij} 
\rho^{\ast}_{i}(\Omega)\rho_{j}(\Omega)
 \label{2.6}
\end{equation}
where

\begin{equation}
\rho_{i}(\Omega)=\sum_{\omega}\phi_{i\downarrow}(-\omega)
\phi_{i\uparrow}(\Omega+\omega)
\label{2.7} 
\end{equation}
with Matsubara's frequencies $\omega=(2m+1)\pi$ and $\Omega=2m\pi$,
($m=0$, $\pm 1$, ..).
In the static approximation, we retain just the term $\Omega=0$ in the 
sum over the frequency $\Omega$. Hence we get for 
$A_{pairing}$  
\begin{equation}
A^{st}_{pairing}=\frac{\beta g}{4 N} \sum_{p=1}^{2}
[\sum_{i\omega}{\underline{\psi}}_{i}^{\dagger}(\omega)
\;\underline{\sigma}_{p}\;\underline{\psi}_{i}(\omega)]^{2}
\label{2.8}
\end{equation}
where we introduced the Nambu matrices
\begin{eqnarray}
{\underline{\psi}}_{i}^{\dagger}(\omega)=
\left(\matrix{\phi^{\ast}_{i\uparrow}(\omega)&
\phi_{i\downarrow}(-\omega)}\right)
\hskip 2cm
\underline{\psi}_{i}(\omega)=
\left(\matrix{\phi_{i\uparrow}(\omega)\cr
	      \phi^{\ast}_{i\downarrow}(-\omega)}\right)
\label{2.9}
\end{eqnarray}
and the Pauli matrices
\begin{eqnarray}
\underline{\sigma}_{1}=\left(\matrix{0&1\cr                         
				     1&0}\right)
\hskip 1cm
\underline{\sigma}_{2}=\left(\matrix{0&-i\cr                         
				     i&0}\right)
\hskip 1cm
\underline{\sigma}_{3}=\left(\matrix{1&0\cr                         
				     0&-1}\right)\;.
\label{2.10}
\end{eqnarray}

The spin part of the action can also be written within the static 
approximation as:
\begin{equation}
A_{SG}=\sum_{ij}\beta J_{ij} S_{i}^{z} S_{j}^{z}
\label{2.11}
\end{equation}
where, from eq. (\ref{2.2}),
\begin{equation}
 S_{i}=\sum_{\omega}{\underline{\psi}}_{i}^{\dagger}(\omega)\; 
\underline{\psi}_{i}(\omega)
\label{2.12}
\end{equation}
Finally, the free action is expressed in terms of Nambu matrices 
\begin{equation}
A_{0}= \sum_{i} {\underline{\psi}}_{i}^{\dagger}(\omega)\;
 \underline{G_{0}}^{-1} (\omega)\;  \underline{\psi}_{i}(\omega)\;.
\label{2.13}
\end{equation}
where the free inverse propagator is
\begin{equation}
\underline{G_{0}^{-1}}(\omega)=\imath\; \omega+
\mu\; \underline{\sigma}_{3}\;.
\label{2.14}
\end{equation}
and the total action can be rebuild as $A=A_{0}+A_{pairing}^{st}+A_{SG}$. 
We are now able to follow the standard procedures to get the 
configurational averaged thermodynamic potential by using the 
replica formalism
\begin{eqnarray}
\Omega=-\frac{1}{\beta} \lim_{n \rightarrow 0} \frac{\langle Z^{n}
\rangle_{ca}-1}{n}
\label{2.15}
\end{eqnarray}
where the configurational averaged replicated partition function, after averaging over $J_{ij}$, 
becomes
\begin{eqnarray}
Z(n)\equiv\langle Z^{n}\rangle_{ca}=
\int D(\phi^{\ast}_{\alpha},\phi_{\alpha})\;
 exp[\sum_{i\alpha\omega} \underline{\psi}^{\dagger\alpha}_{i}(\omega)\;
 \underline{G_{0}^{-1}} (\omega)\;  \underline{\psi}^{\alpha}_{i}(\omega)\nonumber\\+
\frac{\beta g}{4 N}\sum_{\alpha}\sum_{p=1}^{2}\;
[\sum_{i\omega}\underline{\psi}^{\dagger\alpha}_{i}(\omega)\;
\underline{\sigma}_{p}\;
\underline{\psi}^{\alpha}_{i}(\omega)]^{2}\; +\nonumber\\
 \frac{\beta^{2}J^{2}}{2 N}
\sum_{\alpha\beta}[\sum_{i} S^{\alpha}_{i}\; S^{\beta}_{i}]^{2}]
\label{2.16}
\end{eqnarray}
The notation $\underline\psi^{\alpha}_{i}(\omega)$ means that a replica 
index $\alpha=1,2,..,n$ has been associated to each matrix element.
We introduce replica dependent auxiliary fields $\eta_{\alpha}$ and 
$q_{\alpha\beta}$ to linearize the action in eq. (\ref{2.16}), and get 
\begin{eqnarray}
Z(n)=\frac{1}{{\aleph}^n}
\int^{+\infty}_{-\infty}\prod_{\alpha\beta} dq_{\alpha\beta}
\int^{+\infty}_{-\infty}\prod_{\alpha} d\eta_{\alpha R}  
d\eta_{\alpha I} \nonumber\\
e^{-N [\frac{\beta^{2}J^{2}}{2}\sum_{\alpha\beta}q^{2}_{\alpha\beta}
+ \beta g \sum_{\alpha}\eta^{\ast}_{\alpha}\eta_{\alpha}
 -ln\Lambda(q_{\alpha\beta}, \eta_{\alpha})]}
\label{2.17}
\end{eqnarray}
where $\eta_{\alpha}=\eta_{\alpha R}-\imath \eta_{\alpha I}$, 
$\aleph=(\frac{2\pi}{N\beta^{2}J^{2}})(\frac{\pi}{N\beta g})$ and
\begin{eqnarray}
\Lambda(q_{\alpha\beta}, \eta_{\alpha})=\int D(\phi^{\ast}_{\alpha},
\phi_{\alpha}) exp[\sum_{\alpha}\sum_{\omega} 
\underline{\psi}^{\dagger\alpha}(\omega)\; \underline{G_{0}}^{-1} (\omega)
  \;\underline{\psi}^{\alpha}(\omega)+\nonumber\\
\beta g\sum_{\alpha}\sum_{\omega}\underline{\psi}^{\dagger\alpha}(\omega)
\;\underline{\eta}_{\alpha}\; \underline{\psi}^{\alpha}(\omega)+\nonumber\\
\beta^{2}J^{2}\sum_{\alpha\beta} q_{\alpha\beta} S^{\alpha}_{i} 
S^{\beta}_{i}]
\label{2.18}
\end{eqnarray}
while the matrix $\underline{\eta_{\alpha}}$ is defined as 
\begin{equation}
\underline{\eta}_{\alpha}=\left(\matrix{0&\eta_{\alpha}\cr                         
				       \eta^{\ast}_{\alpha}&0}\right)
\label{2.19}
\end{equation}

We analyze the problem within the replica-symmetric ansatz
\begin{equation}
q_{\alpha\neq\beta}=q \hskip 1cm
q_{\alpha\alpha}=q+\overline{\chi} \hskip 1cm
\eta_{\alpha}(\eta^{\ast}_{\alpha})=\eta(\eta^{\ast})
\label{2.20}
\end{equation}
where $q$ is the spin glass order parameter and $\overline{\chi}$ is 
related to the static susceptibility \cite{11} by $\overline{\chi}=\frac{\chi}{\beta}$. 
The complex order parameter $\eta$ gives the number of particle-hole
pairs of opposite spin at each site, as is obtained extremizing $Z(n)$, that is,
 solving $\frac{\partial}{\partial\eta}\langle Z^{n}\rangle=0$ in eq. (\ref{2.17}) and 
the corresponding equation for $\eta^{\ast}$. This yields
\begin{eqnarray}
\eta=\sum_{\omega}\langle\phi^{\ast}_{i\uparrow}(\omega)\phi^{\ast}_{i\downarrow}
(-\omega)\rangle=\langle c^{\dagger}_{i\uparrow} c^{\dagger}_{i\downarrow}\rangle \nonumber\\
\eta^{\ast}=\sum_{\omega}\langle\phi_{i\downarrow}(-\omega)\phi_{i\uparrow}(\omega)
\rangle=\langle c_{i\downarrow} c_{i\uparrow}\rangle\;,
\label{2.21}
\end{eqnarray}
where the brackets indicate both, a statistical average and average over disorder.

The sums over $\alpha$ in the spin part of the action produce again quadratic 
terms that can be linearized by introducing new auxiliary fields, with the result:
\begin{equation}
\Lambda(q,\overline{\chi},\eta)=\int^{+\infty}_{-\infty} dz \frac{ e^{ \frac{-z^{2}}    {2}}}  
 {\sqrt{2\pi}}
\;[\int^{+\infty}_{-\infty} d\xi\; \frac{ e^{ \frac{-\xi^{2}} {2}}}
 {\sqrt{2\pi}}\; I(\xi,z)]^{n}
\label{2.22}
\end{equation}
\begin{eqnarray} 
I(\xi,z)=\int D(\phi^{\ast},\phi) e^{\sum_{\omega} \underline{\psi}^{\dagger}(\omega)\; 
\underline{G}^{-1} (\omega)\;  \underline{\psi}(\omega)}
\label{2.23}
\end{eqnarray}
where the matrix $\underline{G}^{-1}(\omega)$ is given by:
\begin{eqnarray}
\underline{G}^{-1}(\omega)=\left(\matrix{
i\omega+\lambda(z,\xi)+\beta\mu&\beta g \eta\cr                         
	   \beta g \eta^{\ast}        &i\omega+\lambda(z,\xi)-\beta\mu}\right)
\label{2.24}
\end{eqnarray}
and
\begin{equation}
\lambda(z,\xi)=\beta J \sqrt{2 q}\; z+\beta J\sqrt{2 \overline{\chi}}\;\xi\;.
\label{2.25}
\end{equation}
In eq. (\ref{2.23}), the differencial $D(\phi^{\ast},\phi)$ stands for $\prod_{\omega}\prod_{s} d\phi^{\ast}_{s}(\omega)  d\phi_{s}(\omega)$ and the functional integral over Grassmann variables separates into a product of integrals over exponentials of quadratic forms, that can be readily performed with the result \cite{19}:
\begin{eqnarray}
\mbox{ln}I(\xi,z)=\sum_{\omega}\mbox{ln}\mid\underline{G}^{-1}\mid=\nonumber
\sum_{\omega}\mbox{ln}[(\imath\omega+\lambda(z,\xi))^{2}-\nonumber\\
\beta^{2}\mu^{2}-\beta^{2} g^{2}\mid\eta\mid^{2}]\;.
\label{2.26}
\end{eqnarray}

To perform the frequency sum in eq. (\ref{2.26}) one should have in mind that the Nambu formalism introduces a particle-hole transformation in the fermions of spin down. Then from eq. (\ref{2.14}) and eq. (\ref{2.16}) we have that
\begin{equation}
\frac{1}{N}\frac{\partial\Omega}{\partial\mu}=\langle c^{\dagger}_{\uparrow}\;
c_{\uparrow}\rangle-\langle c_{\downarrow}\;c^{\dagger}_{\downarrow}\rangle =\langle n_{\uparrow}\rangle+\langle n_{\downarrow}\rangle-1\;,
\label{2.27}
\end{equation}
and the converging factors in the frequency sums should be adjusted to these prescriptions,
 with the result
\begin{equation}
I(\xi,z) =\mbox{cosh}(\lambda(z,\xi))+\mbox{cosh}(\beta \mu') \;.
\label{2.28}
\end{equation}
where  
\begin{equation}
\mu' =\sqrt{\mu^{2} + g^{2}\eta^{2}}\;.
\label{2.29}
\end{equation}

Not giving rise to confusion, from now on we write $\eta$ in place of$\mid\eta\mid$. 
Introducing eq. (\ref{2.28}) in  eq. (\ref{2.22}) and using eq. (\ref{2.17}) 
we finally obtain for the thermodynamic potential in eq. (\ref{2.15})
at the saddle point:
\begin{eqnarray}
\frac{\beta\Omega}{N}=\frac{1}{2}\beta^{2}J^{2}\;\overline{\chi}\;(2 q+\overline{\chi})
+\beta g \eta^{2}-\nonumber\\
-\int^{+\infty}_{-\infty}
 Dz\; \mbox{ln}
[e^{\beta^{2} J^{2}\; \overline{\chi}} \mbox{cosh}
(\beta J \sqrt{2 q}\; z)+\mbox{cosh}(\beta \mu')]\;.
\label{2.30}
\end{eqnarray}
where $Dz=dz \frac{ e^{ \frac{-z^{2}}    {2}}}   {\sqrt{2\pi}}$. We want, on the average, 
to insure the half-filling situation of one-electron per site, thus fixing $\mu=0$ in 
eqs. (\ref{2.28}) and (\ref{2.30}). The saddle point equations for the order parameters 
that follow from eq. (\ref{2.30}) are:
\begin{equation}
\overline{\chi}=\int Dz\; 
\frac{\mbox{cosh}(\beta J \sqrt{2 q}\; z)}{ \mbox{cosh}(\beta J \sqrt{2 q} \;z)+e^{(-\beta^{2} J^{2}\; 
\overline{\chi})}\;\; \mbox{cosh}(\beta g \eta)}-q
\label{2.32}
\end{equation}
\begin{equation}
q=\int Dz\; 
\frac{\mbox{sinh}^{2}(\beta J \sqrt{2 q}\; z)}{[ \mbox{cosh}(\beta J \sqrt{2 q} 
\;z)+e^{(-\beta^{2} J^{2}\; \overline{\chi})}\;\; \mbox{cosh}(\beta g \eta)]^{2}}
\label{2.33}
\end{equation}
\begin{equation}
\eta=\frac{1}{2} \int Dz\; 
\frac{e^{(-\beta^{2} J^{2} \;\overline{\chi})}\;\; \mbox{sinh}(\beta g \eta)}
{ \mbox{cosh}(\beta J \sqrt{2 q}\; z)+e^{(-\beta^{2} J^{2}\; \overline{\chi})}\;\;
 \mbox{cosh}(\beta g \eta)}\;.
\label{2.34}
\end{equation}

The replica symmetric solution described here is unstable at low temperatures,
 when the de Almeida-Thouless \cite{14} eigenvalue $\lambda_{AT}$ becomes negative.
The calculation of $\lambda_{AT}$ in this model follows as in a  previous work \cite{13}, with the result:
\begin{equation}
\lambda_{AT}=1-\beta^{2}J^{2} \int_{-\infty}^{\infty} Dz\; 
\frac{[1+e^{(-\beta^{2} J^{2} \;\overline{\chi})}\;\; \mbox{cosh}(\beta g \eta)\;\;
\mbox{cosh}(\beta J \sqrt{2 q}\; z)]^{2}}
{[ e^{(-\beta^{2} J^{2} \;\overline{\chi})}\;\; \mbox{cosh}(\beta g \eta)+\mbox{cosh}
(\beta J \sqrt{2 q} \;z)]^{4}}\;.
\label{2.35}
\end{equation}
For the entropy we obtain:
\begin{eqnarray}
\frac{S}{K}=\frac{-3}{2}\beta^{2}J^{2}\;\overline{\chi}\;(2 q+\overline{\chi})
-2 \beta g \eta^{2}+\nonumber\\
\int^{+\infty}_{-\infty}
 Dz \;\;\mbox{ln}
[e^{\beta^{2} J^{2}\; \overline{\chi}}\;\; \mbox{cosh}
(\beta J \sqrt{2 q}\; z)+\mbox{cosh}(\beta g\eta)]\;.
\label{2.36}
\end{eqnarray}
We show in Fig. 5 the behaviour of $\lambda_{AT}$ and $\mbox{S/K}$ as a
function of the temperature for a value of $g>g_{c}$.
 We observe a discontinuity in the derivative of the entropy from the normal to pairing phase 
typical of the second order transition. For lower temperatures $\lambda_{AT}$ and 
$\mbox{S/K}$ become negative
 due to replica symmetry breaking.

A detailed discussion of the numerical solutions of the saddle point equations, as well as the 
Landau expansion of the thermodynamic potential in eq. (\ref{2.30}) in powers of order parameters 
$\mbox{q}$ and $\eta$ is performed in Sec. 3.
\section*{3. Phase Diagram and Tricritical Point}
%

The numerical analysis of the equations for the order parameters $\mbox{q}$, $\eta$ 
and $\overline{\chi}$ in equations (\ref{2.32}), (\ref{2.33}) and (\ref{2.34}) allow 
us to build a phase diagram (temperature versus pairing coupling g) where three regions 
can be identified (see fig.(1)):

i) For high T and small g, we get a normal phase with no long range order where $q=0$ and $\eta=0$.

ii) Enhancing the pairing coupling g, one gets a phase transition at $g=g_{c}(T)$ where there 
is a new order corresponding to the spin pairing on the sites. In terms of the order parameters, 
that means $\eta\neq 0$ while $\mbox{q}=0$.

iii) As one lowers the temperature, for $g<g_{c}(T)$, the model exhibits a phase transition at 
$T=T_{c}$ where $\mbox{q}$ starts to grow continously but with $\eta$ still equal to zero as shown 
in fig. (2). The behaviour of the order parameter $\mbox{q}$ and the  susceptibility 
$\chi=\beta\;\overline{\chi}$ shows a second order transition from a normal phase to a spin glass phase. 
Actually, that situation has been already analysed in \cite{11} where an expansion of equations 
(\ref{2.32}) and (\ref{2.33}) in powers of $\mbox{q}$ for  $\eta=0$ gives $T_{c}=0.9570 J$.

The nature of the transition line given by the equation $g=g_{c}(T)$ is far more complex.
 If $T>T_{c}$, the numerical analysis shows that $\eta$ grows continously from zero as one 
crosses the transition line (see fig. (3)). This result suggests that we get a second order 
transition. Hovewer, when $T<T_{c}$, the numerical solution of the order parameter in fig. (4) 
seems to indicate that the transition line becomes first order at some point. To investigate this
 question further we perform a Landau expansion \cite{22} of the thermodynamic potential 
$\beta\Omega$ in eq. (\ref{2.30}) in powers of the two order parameters $\eta$ and $\mbox{q}$, 
that define the symmetries of the pairing and the spin glass phases, while $\overline{\chi}$ 
is taken at the saddle-point value in eq. (\ref{2.32}). We find it is more convenient to start 
expanding in powers of $\mbox{q}$ and we write from eq. (\ref{2.30}):
\begin{equation}
\beta \Omega=\sum^{3}_{k=0} f_{k}(\eta,\overline{\chi},T)\; q^{k}
\label{3.1}
\end{equation}
where $\overline{\chi}(q,\eta,T)$ is the solution of the saddle point equation
\begin{equation}
\sum^{3}_{k=0}\frac{\partial}{\partial\overline{\chi}} f_{k}(\eta,\overline{\chi},T)\; q^{k}=0\;.
\label{3.2}
\end{equation}
We look for a solution of eq.(\ref{3.2}) also in the form of a series
\begin{equation}
\overline{\chi} =\overline{\chi}_{0}+\overline{\chi}_{1}\; q+\overline{\chi}_{2} \;q^{2}\;.
\label{3.3}
\end{equation}
with the result that $\overline{\chi}_{0}$ is given by
\begin{equation}
\frac{1}{\beta^{2}J^{2}}\;\frac{\partial}{\partial\overline{\chi}} f_{0}(\eta,\overline{\chi}_{0},T)=\overline{\chi}_{0}-\frac{1}{D}=0
\label{3.4}
\end{equation}
where
\begin{equation}
D=1+e^{(-\beta^{2} J^{2}\; \overline{\chi}_{0})}\;\; \mbox{cosh}(\beta g \eta)
\label{3.5}
\end{equation}
and
\begin{equation}
\overline{\chi}_{1}=-[\;(\frac{\partial}{\partial\overline{\chi}}f_{1})\;\;
(\frac{\partial^{2}}
{\partial\overline{\chi}^{2}} 
f_{0})^{-1}\;]_{\eta,\overline{\chi}_{0},T}\;,
\label{3.6}
\end{equation}
\begin{equation}
\overline{\chi}_{2}=-[\;(\frac{\partial}{\partial\overline{\chi}}f_{2}+
\overline{\chi}_{1}\;\frac{\partial^{2}}{\partial\overline{\chi}^{2}} f_{1}+
\frac{1}{2} \overline{\chi}_{1}^{2}\;\frac{\partial^{3}}{\partial\overline{\chi}^{3}}f_{0})\;
(\frac{\partial^{2}}
{\partial\overline{\chi}^{2}} 
f_{0})^{-1})\;]_{\eta,\overline{\chi}_{0},T}\;.
\label{3.7}
\end{equation}
Introducing eq. (\ref{3.3}) into eq. (\ref{3.1}) by expanding the $f_{k}$'s in powers of $\mbox{q}$, we finally obtain after some lengthy calculations the compact result:
\begin{equation}
\beta\Omega=\frac{\beta^{2}J^{2}\overline{\chi}^{2}_{0}}{2}-\mbox{ln}(e^{\beta^{2}J^{2}\overline{\chi}_{0}}+1) + A_{1}\eta^{2}+ A_{2}\eta^{4}+ A_{3}\eta^{6}-B_{1}q^{2}-B_{2}q^{3}
\label{3.8}
\end{equation}
where
\begin{eqnarray}
A_{1}&=&\frac{1}{2!}\;(\beta g)^{2}\;
[\frac{2}{\beta g}
-\frac{1}{\tilde D_{0}}]\nonumber\;,\\ 
A_{2}&=&\frac{1}{4!}\;(\beta g)^{4}\;[\frac{3}{\tilde D^{2}_{0}}-\frac{1}{\tilde D_{0}}]\;,\\
A_{3}&=&\frac{1}{6!}\;(\beta g)^{6}\;[-\frac{1}{\tilde D_{0}}+\frac{15}{\tilde D^{2}_{0}}-\frac{30}{\tilde D^{3}_{0}}]\nonumber\;,\\
B_{1}&=&\beta^{4}J^{4}\;[\frac{1}{2 \beta^{2}J^{2}}-\frac{1}{D^{2}_{0}}]\nonumber\;\\
B_{2}&=&\frac{2}{3}\;\frac{\beta^{6}J^{6}}{D^{3}_{0}}(3 D_{0}+1)\;. 
\label{3.9}
\end{eqnarray}
\begin{eqnarray}
\tilde D_{0}=e^{\beta^{2}J^{2}\;\overline{\chi}_{0}} D_{0}=e^{\beta^{2}J^{2}\;\overline{\chi}_{0}}+1
\label{3.10}
\end{eqnarray}

First we notice that the correct solution of eq. (\ref{3.2}) implies the exact cancellation of the term linear in $\mbox{q}$ in $\beta\Omega$. The order parameter $\eta$ and $\mbox{q}$ minimize and maximize \cite{6}, respectively, $\beta\Omega$ in eq. ( \ref{3.8}). We obtain then that the normal paramagnetic phase is characterized by $A_{1}>0$, $B_{1}>0$; the spin glass phase with $q\neq 0$, $\eta=0$ by  $A_{1}>0$, $B_{1}<0$,; and the pairing phase with $q=0$, $\eta\neq 0$ by $A_{1}<0$, $B_{1}>0$. Lowering the temperature for small values of $\mbox{g}$, $B_{1}$ changes sign first at $T_{c}=0.9570J$ and as $B_{2}>0$ this is a second order transition line from paramagnetic to spin glass phase that was analyzed in detail elsewhere \cite{11}. For $g>g_{c}(T)$ and $T>T_{c}$ we have $A_{1}<0$, $B_{1}>0$, and the line $A_{1}=g-g_{c}(T)=0$ is a second order transition line if $A_{2}>0$. A quick glance at $A_{2}$ shows that it is positive at high temperatures and negative at low temperatures, then we identify the point $g_{3}$, $T_{3}$ where $A_{1}=A_{2}=0$ as a tricritical point, in agreement with the known criteria \cite{13}. At this point, a line of   second order transitions becomes a first order transition line. From eq. (\ref{3.8}) and eq.(\ref{3.4}) we obtain $T_{3}=0.9807J$, $g_{3}=5.8843 J$. The expansion of $A_{1}$ around the tricritical point gives for the critical line:
\begin{equation}
A_{1}=-0.0566(g-g_{3})+0.02575(T-T_{3})=0
\label{3.11}
\end{equation}
and $\eta\approx(T_{P}-T)^{1/2}$ for $T_{3}<T<T_{P}$, where $T_{P}=T_{3}+2.2(g-g_{3})$. For $T<T_{3}$ and $g<g_{3}$, the transition from the pairing to the spin glass phase becomes a first order transition.
Tricritical behaviour has been found previously \cite{16} in the fermionic Ising spin glass model with charge fluctuation, and a discussion of the relation between this model and ours is left for Sec.4.

We show in Fig. 5 the behaviour of $\lambda_{AT}$ and $\mbox{S/K}$ as a function of the temperature for a value of $g>g_{c}$. We observe a discontinuity in the derivative of the entropy from the normal to pairing phase typical of the second order transition. For lower temperatures $\lambda_{AT}$ and $\mbox{S/K}$ become negative due to replica symmetry breaking.

\section*{4. Conclusions}

We study in this paper the interplay of the mechanisms that leads to spin 
glass ordering and BCS pair formation in a soluble mean field model 
Hamiltonian for a fermionic quantum spin glass with a BCS pairing between 
local fermions. As we show in the Appendix, this model would describe the 
spin dynamics of a superconductive spin glass \cite{2,3} and allow us to 
study the role of the pairing mechanism as control of the site occupation 
and the local moment formation. Comparing our results with previous work 
that exhibit tricriticality \cite{15,16}, we can see that the pairing order 
parameter $\eta$ enters in an effective chemical potential $\mu'$ in 
eq. (\ref{2.29}), and our equations would reduce to theirs if we make 
$g=0$, $\mu\neq0$. As we are insuring here half-filling on the average, 
we get $\mu'=g\eta$ and by varying $g$ we change the site occupation by 
favouring doubly occupied states.

As a result we obtain the phase diagram in fig. (1) where we observe a 
normal paramagnetic (NP) phase at high temperatures with $q=\eta=0$. 
By lowering the temperature for $g<g_{c}(T_{c})$ we encounter a second 
order transition line from the NP phase to the SG (spin glass) phase at 
$T_{c}=0.9570J$.
For $g>g_{c}(T_{c})$ the second order transition is from the NP phase to 
the PAIR ( pairing formation) phase if $T_{3}<T<T_{P}(g)$. At $T=T_{3}$, 
$g=g_{3}$ there is a tricritical point where the pairing transition becomes 
first order. This point is almost indistinguishable in the fig. (1) from the 
point $T=T_{c}$, $g=g_{c}$. For $T<T_{c}$ the line $g_{c}(T)$ becomes a 
first order transition line separating the SG and PAIR phases. The phase 
diagram obtained in fig. (1) is in good qualitative agreement with the 
experimental results of ref. \cite{2} for 
$\mbox{Gd}_{x}\mbox{Th}_{1-x}\mbox{RU}_{2}$ samples, by assuming that 
the ratio $J/g$ is proportional to the $\mbox{Gd}$ concentration.
This assumption is reasonable, as we show in the Appendix that the effective
value of g is proportional to the number of superconductive pairs, that
for a given temperature decreases drastically with the concentration of
magnetic impurities, leading to an increase in $J/g$.
To conclude, we studied a model for a fermionic SK spin glass with BCS 
pairing among the local fermions that is  soluble by reduction to a
one site problem. Although this model originates in the description of the 
spin dynamics of conventional spin glass superconductors, we hope that 
these results may be also relevant for the study of 
strongly correlated fermions 
systems through the localized one site approximations \cite{23}. 
It is possible to extend the analysis of Ref.\cite{10} for
the study of the time correlation functions in the present problem,
but it will be the subject of a future work.

\section*{Acknowledgements}
We are grateful to W. K. Theumann for relevant comments and we thank 
P. Pureur
for discussions. S. G. Magalh\~aes
acknowledges the hospitality of the Instituto de F\'{\i}sica,
UFRGS, where part of this work was performed. This work was partially 
supported by CNPq (Conselho Nacional de Desenvolvimento Cient\'{\i}fico e 
Tecnol\'ogico) and FINEP (Financiadora de Estudos e Projetos).

\newpage
\section*{Appendix}

Conventional spin glass superconductors are usually represented by a 
system of conduction electrons with BCS coupling interacting with 
localized spins \cite{3}. Using Gorkov's decoupling scheme the 
Hamiltonian is:
\begin{eqnarray}
\overline{H}_{alloy}= \sum_{ks}(\epsilon_{k}-\mu) a^{\dagger}_{ks}a_{ks}- 
 \sum_{k}[\Delta_{k}a^{\dagger}_{k\uparrow}a^{\dagger}_{-k\downarrow}+
\Delta^{\dagger}_{k}a_{-k\downarrow}a_{k\uparrow}]\nonumber\\
-J_{sd}\;\sum_{i}\vec{S}_{i}.\vec{s}_{i}
\label{A.1}
\end{eqnarray}
where $\vec{S}_{i}$ is the magnetic moment localized at the random site 
$\vec{R}_{i}$ and $\vec{s}_{i}$ is the local spin density of the conduction 
electrons
\begin{equation}
\vec{s}_{i}= \sum_{k k'}\sum_{s s'} e^{i(\vec{k}-\vec{k}').\vec{R}_{i}}
a^{\dagger}_{ks}\vec{\sigma}_{s s'}a_{k' s'}
\label{A.2}\;,
\end{equation}
where $\vec{\sigma}_{s s'}$ indicates the elements of the vector 
Pauli matrices and $a^{\dagger}_{ks} (a_{k s}$) are the usual creation
 (annihilation) operators for superconducting electrons. 
The order parameters $\Delta_{k}$, $\Delta^{\dagger}_{k}$ are to be 
determined self-consistently from the equations of motion,
but here we consider them to be phenomenological parameters. 
In this paper we choose to represent the localized moments by 
a bilinear combination of fermion operators \cite{12}
\begin{equation}
\vec{S}_{i}=\sum_{s s'} c^{\dagger}_{is}\vec{\sigma}_{s s'}c_{i s'}
\label{A.3}
\end{equation}
as we did in eq. (\ref{2.2}).

The partition function for the superconducting alloy may be written in 
terms of functional integrals as we did in eq. (\ref{2.4}) and 
eq. (\ref{2.5}): 
\begin{equation}
Z_{alloy}=
\int\prod_{is} D(\phi^{\ast}_{is},\phi_{is}) 
\int\prod_{ks} D(\phi^{\ast}_{ks},\phi_{ks})\; e^{A_{alloy}}
\label{A.4}
\end{equation}
where the action is now given by
\begin{eqnarray}
A_{alloy}=\int_{0}^{\beta}
[\sum_{is}\phi^{\ast}_{is}(\tau)\frac{d}{d\tau}\phi_{is}(\tau)     + \sum_{ks}\phi^{\ast}_{ks}(\tau)\frac{d}{d\tau}\phi_{ks}(\tau)\nonumber\\
-H_{alloy}(\phi^{\ast}(\tau),\phi (\tau)]d\tau
\label{A.5}
\end{eqnarray}
and we introduced the Grassmann fields $\phi^{\ast}_{ks}$, $\phi_{ks}$ 
for the conducting band. Using the Nambu formalism as we did
 in Sec. 2 we may write the partition function
\begin{equation}
Z_{alloy}=
\int\prod_{is} D(\phi^{\ast}_{is},\phi_{is})\; e^{A_{0}}
\int\prod_{ks} D(\phi^{\ast}_{ks},\phi_{ks})\; e^{A_{BCS}+A_{sd}}
\label{A.6}
\end{equation}
where $\mbox{A}_{0}$ is the action for non-interacting fermions given 
in eq. (\ref{2.13}) and $\mbox{A}_{BCS}$ is the action for the 
superconducting electrons:
\begin{equation}
A_{BCS}= \sum_{k\omega} {\underline{\psi}}_{k}^{\dagger}(\omega)\;
 \underline{G_{k}}^{-1} (\omega)\;  \underline{\psi}_{k}(\omega)\;.
\label{A.7}
\end{equation}
where similarly to eq. (\ref{2.9}) and eq. (\ref{2.24})

\begin{eqnarray}
{\underline{\psi}}_{k}^{\dagger}(\omega)=
\left(\matrix{\phi^{\ast}_{k\uparrow}(\omega)&
\phi_{-k\downarrow}(-\omega)}\right)
\hskip 2cm
\underline{\psi}_{k}(\omega)=
\left(\matrix{\phi_{k\uparrow}(\omega)\cr
	      \phi^{\ast}_{-k\downarrow}(-\omega)}\right)
\label{A.8}
\end{eqnarray}
\begin{eqnarray}
\underline{G}^{-1}_{k}(\omega)=\left(\matrix{
i\omega-\beta(\epsilon_{k}-\mu)&\beta\Delta_{k}\cr                         
	   \beta \Delta^{\ast}_{k}      &i\omega+\beta(\epsilon_{k}-\mu)}\right)
\label{A.9}
\end{eqnarray}
The s-d exchange part of the action is given by
\begin{equation}
A_{sd}=-J_{sd}\beta\sum_{i}\sum_{\Omega} \vec{S}_{i}(-\Omega).\vec{s}_{i}(\Omega)
\label{A.10}
\end{equation}
where from eq. (\ref{A.2}) and eq. (\ref{A.3})
\begin{equation}%
\vec{S}_{i}(\Omega)=\sum_{s s'} \sum_{\omega} \phi^{\ast}_{is}(\omega+\Omega)\vec{\sigma}_{s s'}\phi_{i s'}(\omega)
\label{A.11}
\end{equation}
\begin{equation}
\vec{s}_{i}(\Omega)=\sum_{s s'}\sum_{k k'} \sum_{\omega} e^{i(\vec{k}-\vec{k}').\vec{R}_{i}} \phi^{\ast}_{ks}(\omega+\Omega)\vec{\sigma}_{s s'}\phi_{k's'}(\omega)
\label{A.12}
\end{equation}

We indicate by $\omega=(2n+1)\pi$ and $\Omega=2n\pi$ the 
fermionic and  bosonic Matsubara frequencies, respectively. 
When we are interested in the localized spins dynamics, the 
conduction electrons degrees of freedom in eq. (\ref{A.6}) may be 
integrated out to second order perturbation theory in $\mbox{J}_{sd}$
to give the result:
\begin{equation}
Z_{alloy}=
\int\prod_{is} D(\phi^{\ast}_{is},\phi_{is})\; e^{A_{0}+A_{eff}(\phi^{\ast}_{is},\phi_{is})}
\label{A.13}
\end{equation}
where
\begin{eqnarray}
 A_{eff}(\phi^{\ast}_{is},\phi_{is})=-\frac{1}{2}(\beta J_{sd})^{2}
\sum_{\Omega\alpha\beta}\sum_{ij}V^{\alpha\beta}_{ij}(\Omega)S^{\alpha}_{i}
(\Omega)S^{\beta}_{j}(-\Omega)\nonumber\\
-\frac{1}{2}(\beta J_{sd})^{2}
\sum_{\Omega}\sum_{ij}W_{ij}(\Omega)
[S^{+}_{i}(\Omega)S^{-}_{j}(\Omega)+S^{-}_{i}(\Omega)S^{+}_{j}(-\Omega)]
\label{A.14}
\end{eqnarray}
and the dynamic interactions $V^{\alpha\beta}_{ij}(\Omega)$, $W_{ij}(\Omega)$ are obtained from the correlation functions $\langle S^{\alpha}_{i}(\Omega)S^{\beta}_{j}(-\Omega) \rangle$. By performing the frequency sums as indicated in Sec. 2 and approximating $\Delta_{k}\approx \Delta$, we obtain for the static part with $\Omega=0$:
\begin{equation}
V^{\alpha\beta}_{ij}(0)=
\delta_{\alpha\beta}\sum_{\vec{q}}
e^{i\vec{q}.(\vec{R}_{i}-\vec{R}_{j})}\frac{1}{\beta} 
\sum_{\vec{k}}\frac{n(\vec{k}+\vec{q})-n(\vec{k})}{\epsilon(\vec{k}+\vec{q})-
\epsilon(\vec{k})}
\label{A.15}
\end{equation}
\begin{equation}
W_{ij}(0)=\frac{|\Delta|}{\beta}\sum_{\vec{q}}
e^{i\vec{q}.(\vec{R}_{i}-\vec{R}_{j})} 
\sum_{\vec{k}}\frac{B(\vec{k})-B(\vec{k}+\vec{q})}{E^{2}(\vec{k}+\vec{q})-
E^{2}(\vec{k})}
\label{A.16}
\end{equation}
where
\begin{eqnarray}
n(\vec{q})=[1+e^{\beta(\epsilon(\vec{q})-\mu)}]^{-1} \nonumber\\
B(\vec{k})=\frac{|\Delta|}{2 E(\vec{k})}tanh(\frac{\beta E(\vec{k})}{2})\\
E(\vec{k})=[(\epsilon(\vec{k})-\mu)^{2}+
|\Delta|^{2}]^{\frac{1}{2}}\nonumber
\label{A.17}
\end{eqnarray}

The effective interaction $V_{ij}$ in eq. (\ref{A.15}) was calculated to lowest order in $|\Delta|$ and it represents the familiar RKKY interaction \cite{24} that is responsible for spin glass ordering. The interaction $W_{ij}$ is of a different character and it represents the coupling induced by the exchange of superconducting electrons. The function $B(\vec{k})$ is the matrix element \cite{25} of the pairing operator $c^{\dagger}_{k\uparrow}c^{\dagger}_{-k\downarrow}$, and since pairing interactions smooth out the jump in the single particle occupation number $n(\vec{k})$ we can approximate $B(\vec{k})=\frac{|\Delta|}{2E(\vec{k})}$ at low temperatures, what gives in eq. (\ref{A.16}).
\begin{equation}
W_{ij}(0)\approx\frac{1}{2\beta}\sum_{\vec{q}}
e^{i\vec{q}.(\vec{R}_{i}-\vec{R}_{j})} 
\sum_{\vec{k}}\frac{|\Delta|^{2}}{E(\vec{k}+\vec{q})E(\vec{k})}
[\frac{1}{E(\vec{k}+\vec{q})+E(\vec{k})}]
\label{A.18}
\end{equation}
As the last sum in eq. (\ref{A.18}) is very weakly dependent on $\vec{q}$ it can be approximated by its value when $q=0$, what gives
\begin{equation}
W_{ij}(0)\approx\frac{1}{4\beta}\sum_{\vec{k}}
\frac{|\Delta|^{2}}{E^{3}(\vec{k})} \delta_{ij}
\label{A.19}
\end{equation}

Introducing eq. (\ref{A.15}) and eq. (\ref{A.19}) in eq. (\ref{A.14}) we obtain for the static part of the interaction
\begin{equation}
 A^{st}_{eff}(\phi^{\ast}_{is},\phi_{is})\approx-
\sum_{ij}J_{ij}(RKKY)\vec{S}_{i}(0).\vec{S}_{j}(0)-
g \sum_{i} S^{+}_{i}(0)S^{-}_{i}(0)
\label{A.20}
\end{equation}

We argue that the last term in eq. (\ref{A.20}) is the static part of the action corresponding to the Hamiltonian $H_{I}=g \sum_{i}c^{\dagger}_{i\uparrow}c_{i\downarrow}c^{\dagger}_{i\downarrow}
c_{i\uparrow}$ that within Gorkov's formalism would give rise to terms $-g [\Delta_{l}^{\dagger} \sum_{i}c_{i\downarrow}c_{i\uparrow}+\Delta_{l}\sum_{i}c^{\dagger}_{i\uparrow}
c^{\dagger}_{i\downarrow}]$, what ultimately justifies our choice of Hamiltonian in eq. (\ref{2.1}).

\newpage

%
%
\newpage
\section*{Figure captions}
\noindent
{\bf Figure 1:} Phase diagram as function of temperature and pairing coupling $\mbox{g}/\mbox{J}$. Solid lines indicate second order transitions while the dotted line indicates a first order transition. The tricritical point $\mbox{T}_{3}$, $\mbox{g}_{3}$ is shown in detail in the diagram. The points where $\lambda_{AT}$ becomes negative are represented by the dashed line. \\
\noindent
{\bf Figure 2:} Temperature behaviour of $\mbox{q}$ (solid line), and $\overline{\chi}=\frac{\chi}{\beta}$ (dashed line)
for $\mbox{g}=0.5\mbox{J}< \mbox{g}_{c}$. Here $\eta=0$.\\
\noindent
{\bf Figure 3:} Dependence of the order parameter $\eta$ (dotted line) and the parameter $\overline{\chi}$ (dashed line) with the coupling $\mbox{g}/\mbox{J}$ for $\mbox{T}>\mbox{T}_{c}$, where $\mbox{T}=1.5J$. We show in detail the continuous behaviour of $\eta$ around $g_{c}$.\\
\noindent
{\bf Figure 4:} Dependence of the order parameters $\eta$ (dotted line),
$\mbox{q}$ (solid line), and the parameter $\overline{\chi}$ (dashed line) with the coupling  $\mbox{g}/\mbox{J}$ for $\mbox{T}<\mbox{T}_{c}$, where $\mbox{T}=0.75J$. At the transition point both $\eta$ and $q$ have discontinous behaviour indicating a first order transition. \\
\noindent
{\bf Figure 5:} The Almeida-Thouless eigenvalue (solid line)
and entropy (dashed line) for $\mbox{g}=6.2\mbox{J}$.\\
%
\\

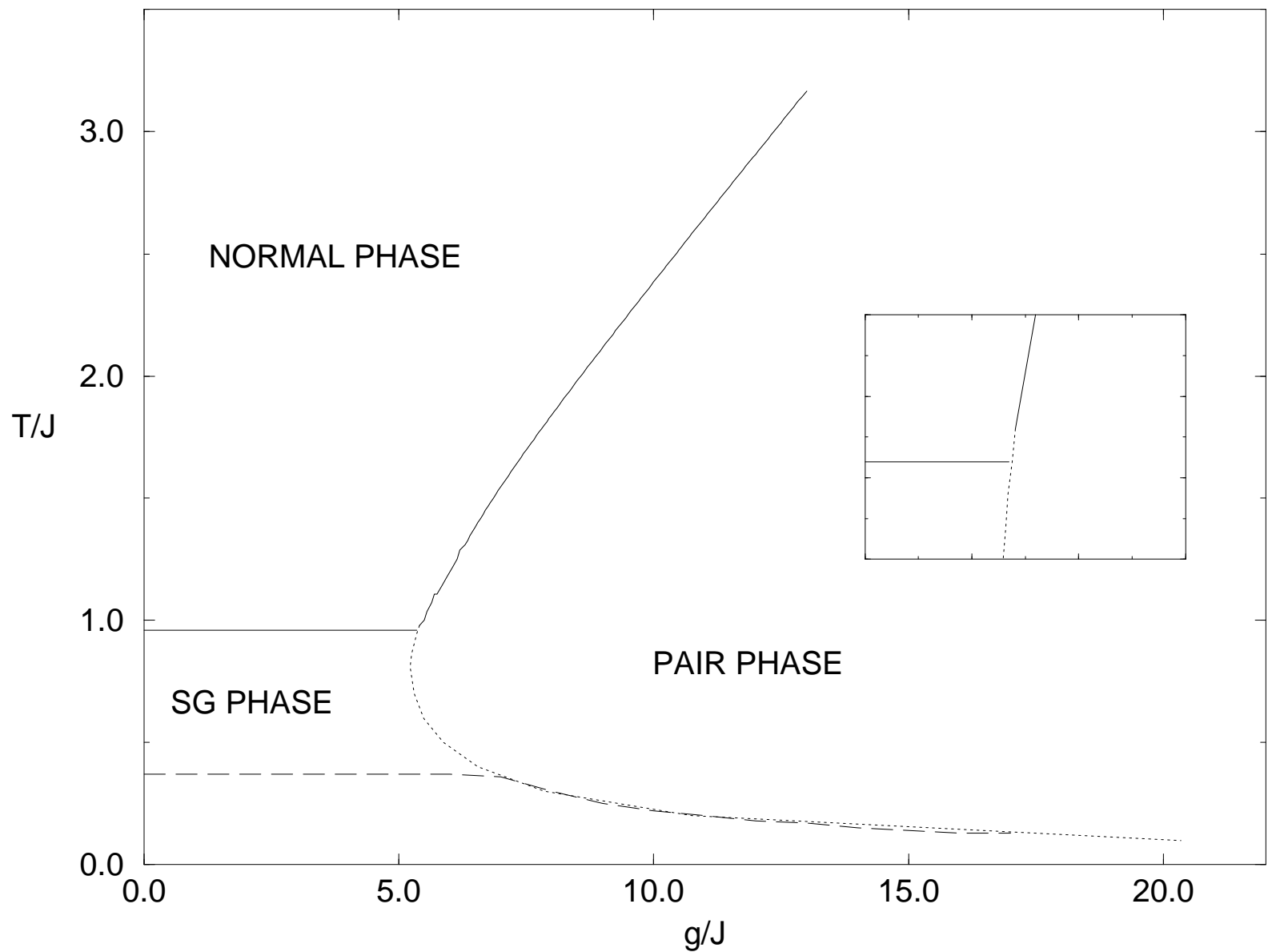
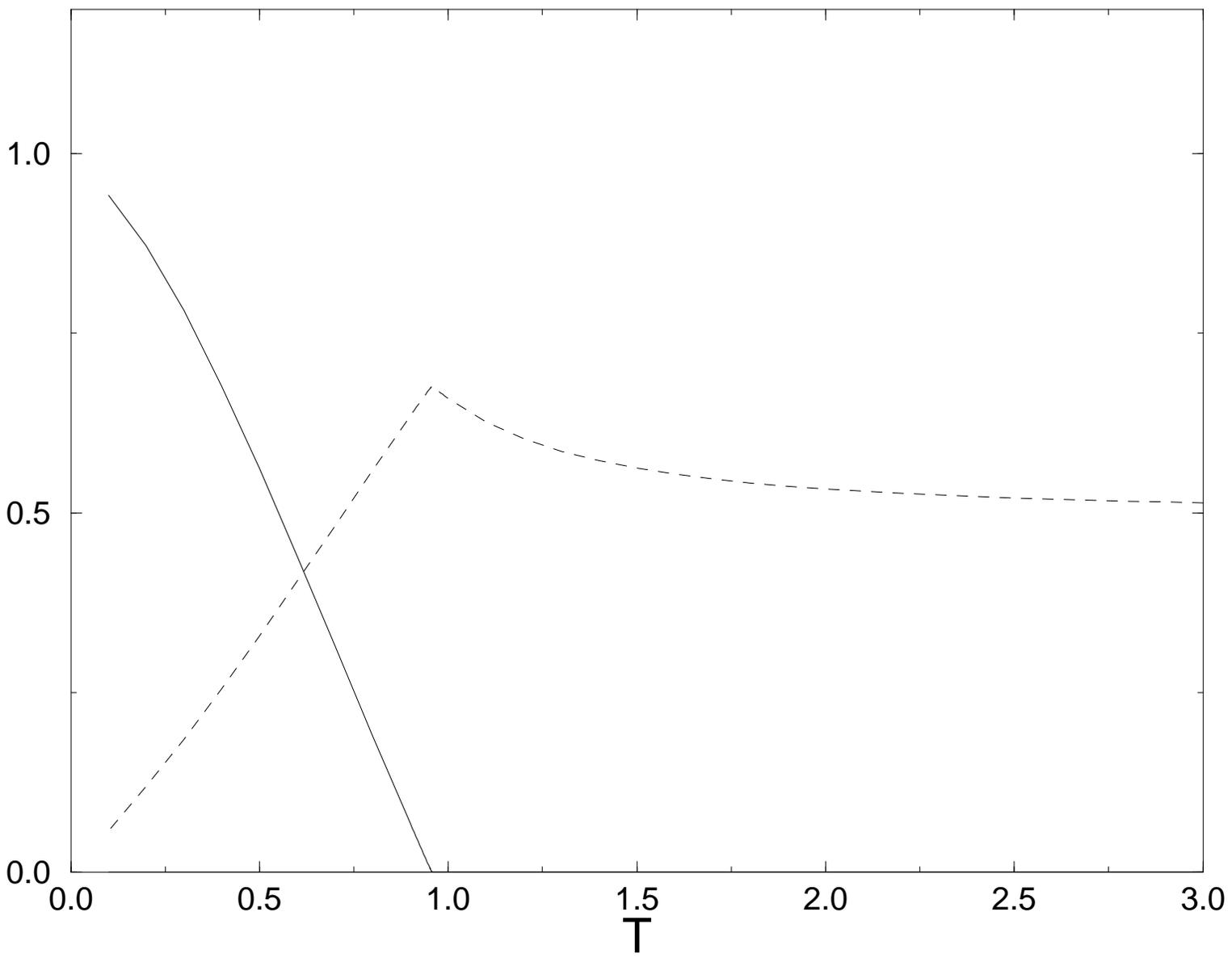
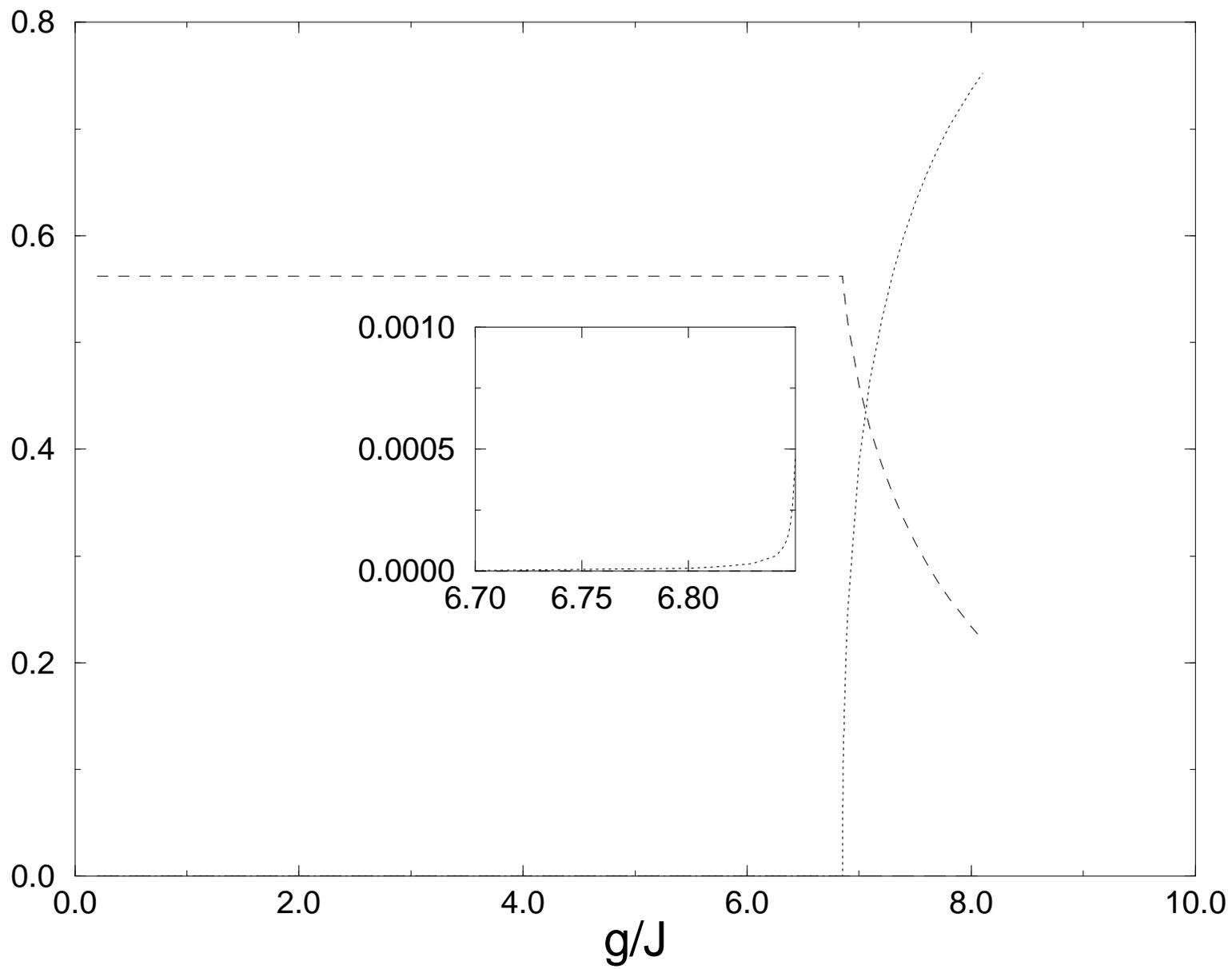
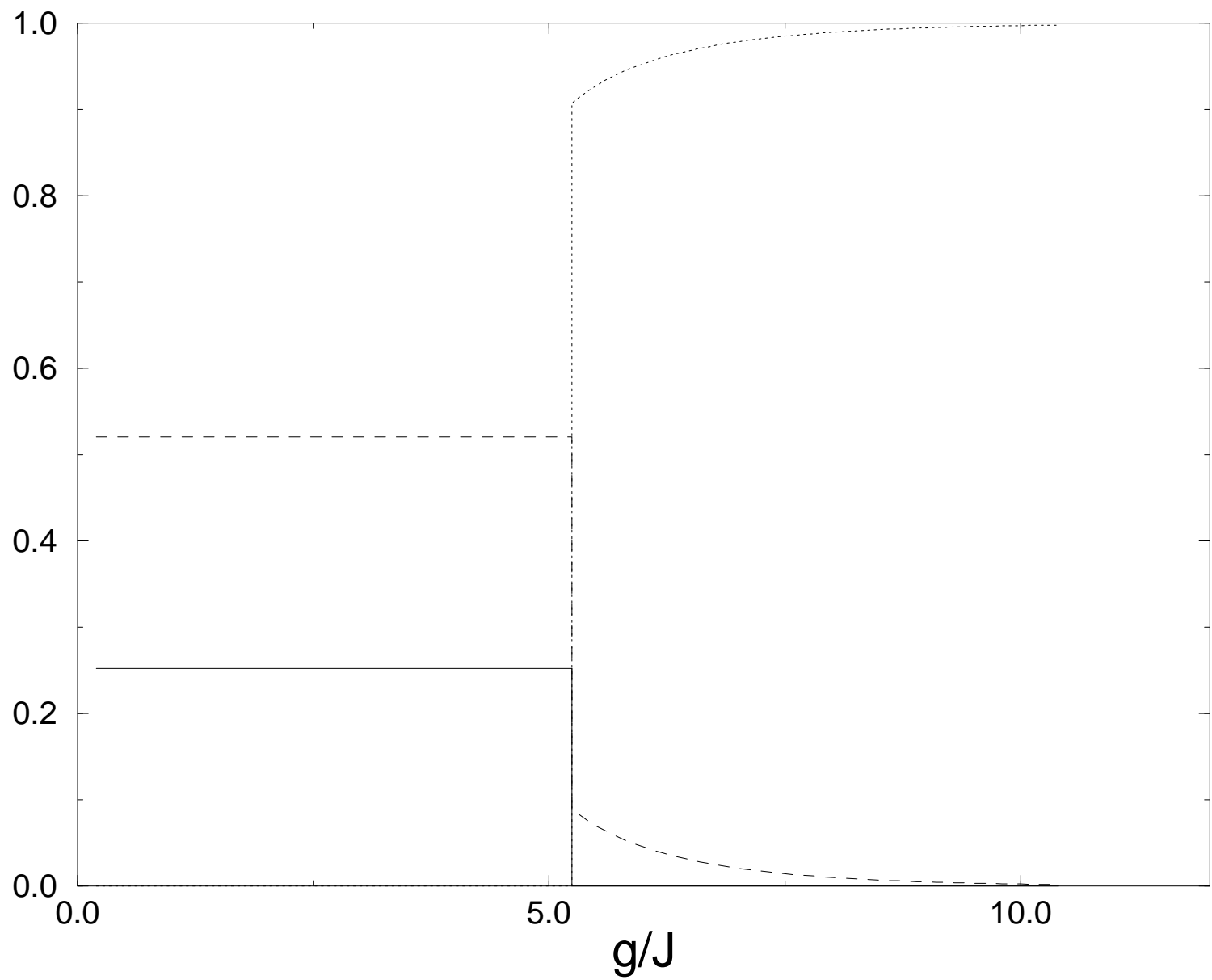
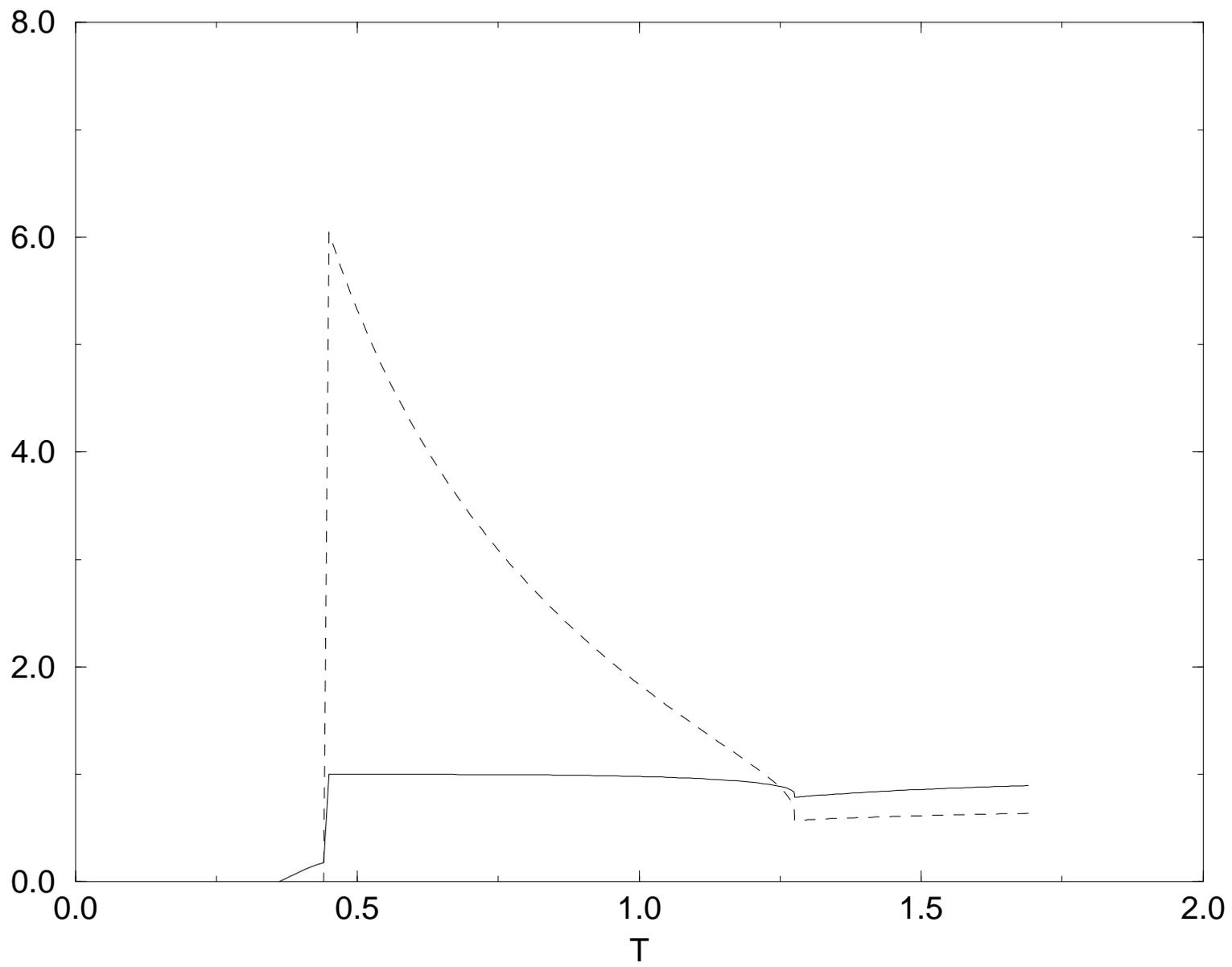
\end{document}